\documentclass[a4paper]{svmult}
\usepackage{times}
\usepackage[dvips]{graphicx,epsfig}
\usepackage{amsmath,amsfonts,amssymb}
\usepackage{cite,url}

\begin{document}



\mainmatter

\newcommand{\be}{\begin{equation}} \newcommand{\ee}{\end{equation}}
\newcommand{\ba}{\begin{eqnarray}} \newcommand{\ea}{\end{eqnarray}}
\newcommand{\uu}{u({\bf p},s)} \newcommand{\ubar}{\overline{u}({\bf
    p'},s')} \newcommand{\nk}{{\bf k}} \newcommand{\nq}{{\bf q}}
\newcommand{\np}{{\bf p}} \newcommand{\nh}{{\bf h}}
\newcommand{\nO}{{\bf 0}} \newcommand{\nkappa}{\mbox{\boldmath
    $\kappa$}} \newcommand{\nx}{{\bf x}}
\newcommand{\neta}{\mbox{\boldmath $\eta$}} 
\newcommand{\nsigma}{\mbox{\boldmath $\sigma$}} 
\newcommand{\ny}{{\bf y}}
\newcommand{\nR}{{\bf R}} \newcommand{\sigvec}{\mbox{\boldmath
    $\sigma$}} \newcommand{\etavec}{\mbox{\boldmath $\eta$}}
\newcommand{\tauvec}{\mbox{\boldmath $\tau$}}
\newcommand{\kappavec}{\mbox{\boldmath $\kappa$}}
\newcommand{\kvec}{\mbox{\boldmath $\zeta$}}
\newcommand{\avec}{\mbox{\boldmath $A$}}
\newcommand{\bvec}{\mbox{\boldmath $B$}}
\newcommand{\gammavec}{\mbox{\boldmath $\gamma$}}
\newcommand{\Gammavec}{\mbox{\boldmath $\Gamma$}}
\newcommand{\jvec}{\mbox{\boldmath $J$}}
\newcommand{\hvec}{\mbox{\boldmath $h$}} \newcommand{\bd}[1]{
  \mbox{\boldmath $#1$} } \newcommand{\sla}[1]{#1 \!\!\!\!\!\!\slash }
\def\be{\begin{equation}}
\def\ee{\end{equation}}
\def\bea{\begin{eqnarray}}
\def\eea{\end{eqnarray}}
\def\bear{\begin{array}}
\def\ear{\end{array}}
\def\bfig{\begin{figure}}
\def\efig{\end{figure}}
\def\bcen{\begin{center}}
\def\ecen{\end{center}}
\def\raw{\rightarrow}
\def\bra#1{\left\langle #1\right|}
\def\ket#1{\left| #1\right\rangle}
\def\Bra#1{\bigl\langle #1\bigr|}
\def\Ket#1{\bigl| #1\bigr\rangle}
\def\vp{\mathbf{p}}
\def\vk{\mathbf{k}}
\def\vv{\mathbf{v}}
\def\vq{\mathbf{q}}
\def\mpi{m_{\pi}}
\def\la{\label}
\def\chic{\scriptscriptstyle}
\def\S{\scriptstyle}
\def\D{\displaystyle}
\def\slash{\!\!\! /} 
\def\bpi{\boldsymbol\pi}
\def\btau{\boldsymbol\tau}
\def\bsigma{\boldsymbol\sigma}
\def\bphi{\boldsymbol\phi}
\def\brho{\boldsymbol\rho}
\def\bnu{\boldsymbol\nu}
\def\bmu{\boldsymbol\mu}
\def\bkappa{\boldsymbol\kappa}
\def\boeta{\boldsymbol\eta}

\def\three_j(#1,#2,#3,#4,#5,#6){\pmatrix{#1 & #2 & #3\cr
                                         #4 & #5 & #6\cr}}

\def\qqq{\end{document}}
\def\pmb#1{\setbox0=\hbox{$#1$}%
\kern-.025em\copy0\kern-\wd0
\kern.05em\copy0\kern-\wd0
\kern-.025em\raise.0433em\box0 }
\def\b{\pmb}

\def\xara(#1,#2,#3,#4){\left(\matrix{#1 & #2\cr #3 & #4\cr}\right)}
\def\Fn{J_{(n)}}
\def\Fs{J_{(s)}}
\def\Hn{H_{(n)}}
\def\Hs{H_{(s)}}
\def\thru#1{\mathrel{\mathop{#1\!\!\!/}}}
\def\CN{\cal N}
\def\w{\omega}
\def\W{\Omega}
\def\six_j(#1,#2,#3,#4,#5,#6){\left\{\matrix{#1 & #2 & #3\cr
                                         #4 & #5 & #6\cr}\right\}}
\def\nine_j(#1,#2,#3,#4,#5,#6,#7,#8,#9){\left\{\matrix{#1 & #2 & #3\cr
                                        #4 & #5 & #6\cr
                                         #7 & #8 & #9\cr}\right\}}
\def\W{\Omega}
\def\pd#1#2{{\partial #1\over \partial #2}}
\def\v#1{ {\bf #1} }
\def\Ener(#1,#2){ \sqrt{{#1}^2+{#2}^2} }
\def\c#1{ {\cal #1}}%

\def\overlay#1#2{\setbox0=\hbox{$#1$}\setbox1=\hbox to \wd0{\hss$#2$\hss}#1%
\hskip -1\wd0\copy1}
\newcommand{\xslash}[1]{\overlay{#1}{/}}
\newcommand{\undsim}[1]{\olay{#1}{\sim}}

\def\bold#1{\setbox0=\hbox{$#1$}%
      \kern-.025em\copy0\kern-\wd0
      \kern.05em\copy0\kern-\wd0
      \kern-.025em\raise.0433em\box0 }
\def\tr{\, \hbox{tr} \, }
\def\Tr{\, \hbox{Tr} \, }
\def\sgn{\, \hbox{sgn} }
\def\bra{\langle}
\def\ket{\rangle}
\def\shalf{\,1/2\,}
\def\half{\, {1 \over 2} \,}
\def\gsim{\displaystyle\mathop{>}_{\sim}}
\def\lsim{\displaystyle\mathop{<}_{\sim}}
\def\rd{\partial}
\def\c{\hbox{c}}
\def\s{\hbox{s}}

\def\S11{S_{11}(1535)}
\def\etaNN{\eta NN^*}
\def\geta{g_{\eta NN^*}}

\def\tdotr{\, \vec \tau \cdot \hat r \, }
\def\fpi{f_\pi}

\renewcommand{\thefootnote}{\fnsymbol{footnote}}
\def\footnoterule{\kern-3pt \hrule width \hsize \kern2.6pt}

\newcommand{\nne}{{\bf e}}
\newcommand{\im}{{\rm Im}\,}
\newcommand{\re}{{\rm Re}\,}
\newcommand{\qbar}{\not{\!Q}}
\newcommand{\pbar}{\not{\!P}}
\newcommand{\pdbar}{\not{\!P}_{\!\Delta}}

%
\newcommand{\md}{m_\Delta}

\newcommand{\na}{{\bf      a}}  
\newcommand{\nb}{{\bf      b}}     
\newcommand{\nj}{{\bf      j}}
\newcommand{\nr}{{\bf      r}}         
\newcommand{\ns}{{\bf      s}}
\newcommand{\nv}{{\bf      v}}
\newcommand{\nw}{{\bf      w}}
\newcommand{\nA}{{\bf      A}}       
\newcommand{\nB}{{\bf      B}}
\newcommand{\nC}{{\bf      C}}
\newcommand{\nJ}{{\bf      J}}
\newcommand{\nM}{{\bf      M}}       
\newcommand{\nP}{{\bf      P}}       
\newcommand{\nS}{{\bf      S}}      
\newcommand{\nX}{{\bf      X}}
\newcommand{\nY}{{\bf      Y}}          
\newcommand{\hp}{{\bf \hat{p}}}
\newcommand{\hr}{{\bf \hat{r}}} 
\newcommand{\hx}{{\bf \hat{x}}} 

\def\M{M^{0,(RFG)}_A}
\def\MA{(M^0_{A-1})^2}
\def\E{{\cal E}}
\def\N{N_{\rm spin}^{\rm isospin}}
\def\up{u^\prime}
\def\vp{v^\prime}
\def\upsq{u^{\prime\, 2}}
\def\vpsq{v^{\prime\, 2}}

\title*{A simple model for NN correlations in quasielastic lepton-nucleus 
scattering}

\author{\underline{{M.B. Barbaro}}\inst{1}
\and {R. Cenni}\inst{2}
\and {T.W. Donnelly}\inst{3}
\and {A. Molinari}\inst{1}
}

\titlerunning{A simple model for NN correlations in quasielastic 
lepton-nucleus scattering}
\authorrunning{{Maria B. Barbaro} {\em et al.}}

\toctitle{A simple model for NN correlations in quasielastic 
lepton-nucleus scattering}
\tocauthor{{M.B. Barbaro}, {R. Cenni}, {T.W. Donnelly}, {A. Molinari}
}

\institute{
{Universit\`a di Torino and INFN, Sezione di Torino, Italy}
\and
{INFN, Sezione di Genova, Italy}
\and
{CTP, LNS and Department of Physics, MIT, 
  Cambridge, USA}
}

\maketitle

\begin{abstract}

We present a covariant extension of the relativistic Fermi gas model
which incorporates  correlation effects in nuclei.
Within this model, inspired by the BCS descriptions of systems of fermions,
we obtain the nuclear spectral function and from
it the superscaling function for use in treating high-energy
quasielastic electroweak processes. Interestingly, this model has
the capability to yield the asymmetric tail seen in the experimental
scaling function.
\end{abstract}

\section{Introduction}

Recently the theoretical understanding of quasielastic (QE) lepton-nucleus 
scattering has received renewed attention not only
because of its intrinsic interest but also because reliable calculations
of neutrino-nucleus cross section in the QE domain are essential 
when addressing fundamental neutrino properties, specifically neutrino masses 
and the neutrino oscillations that result from those masses. 

In particular it has been suggested~\cite{nu}
that superscaling~\cite{Day,DS} in electroweak 
interactions with nuclei, namely the observation that the reduced
electron-nucleus cross sections are to a large degree independent of the
momentum transfer (scaling of I kind) and of the nuclear species (scaling of
II kind), can be used as a tool to obtain predictions for
neutrino-nucleus cross sections. 
Owing to the complexity of nuclear dynamics it is not obvious that
the nuclear response to an electroweak field superscales. 
Indeed several effects are expected to break superscaling to some extent:
off-shellness, collective nuclear excitations, meson-exchange 
currents, nucleon-nucleon (NN) correlations. To assess the impact of these 
contributions in the QE peak region is then of crucial 
importance.

In the present work we explore the role of NN correlations in the QE peak 
domain proposing an extension of the Relativistic Fermi Gas (RFG) approach
which still  includes only on-shell nucleons in an independent-particle
model, and put our efforts into going
beyond the degenerate description provided by the extreme RFG.
To do this we resort to a model~\cite{Barbaro:2008zv} 
inspired by the BCS theory of
condensed matter physics with appropriate modifications, such as retention
of covariance, to adapt it to the high-energy physics of atomic nuclei.


\section{Longitudinal response and superscaling function}
\label{sec:resp}

We concentrate here on the
longitudinal electromagnetic nuclear response $R_L$, namely the
part of the total inclusive electroweak response that is believed to
superscale the best~\cite{DS}. All other electroweak responses can be
developed using similar arguments to those presented in the
following. 
Within the framework of the plane-wave impulse approximation, 
where it is assumed that only one vector boson is
exchanged between the probe and the nucleus and that this one is
absorbed by a single nucleon,  the QE
longitudinal response function of 
a nucleus to an external electroweak field bringing three-momentum $\bf q$ and
energy $\omega$ into the system reads
\begin{equation}
\label{eq:RL} 
R_L(q,\omega)={\cal N}\, R_L^{s.n.}(q,\omega)\, \frac{2\pi m_N^2}{q}
\int\!\!\!\!\int_\Sigma dp\,d{\cal E}\,\frac{p}{E_p} S(p,{\cal E})\, ,
\end{equation}
where ${\cal N}$ is the appropriate nucleon number ($Z$ for protons and $N$
for neutrons), $R_L^{s.n.}(q,\omega)$ is the corresponding single
nucleon response and $E_p=\sqrt{p^2+m_N^2}$ 
is the on-shell energy of the struck nucleon, with $m_N$ the nucleon mass.
The probability of finding one
nucleon in the system is provided by the system's spectral function 
$S(p,{\cal E})$,
which depends on the missing momentum $p$ and on the energy 
\begin{equation}
{\cal E} = \sqrt{p^2+M^{* 2}_{A-1}}-\sqrt{p^2+M^{2}_{A-1}}
= \omega-T_N-E_s-T_{A-1}\,.
\end{equation}
The latter is the excitation energy of the residual nucleus in the reference 
frame where it moves with momentum $-{\bf p}$ and, neglecting the very small recoiling nucleus kinetic energy $T_{A-1}$,
is essentially the missing energy $\omega-T_N$, $T_N$ being the ejected 
nucleon kinetic energy, minus the separation energy $E_s=M_{A-1}+m_N-M_A$.

Equation~\eqref{eq:RL} connects the semi-inclusive $(l,l'N)$
reaction with the inclusive $(l,l')$ process assuming that the
outgoing nucleon no longer interacts with the residual $(A-1)$
nucleus (absence of final-state interactions). That equation
expresses the assumption that the inclusive cross section is to be
obtained by integrating the semi-inclusive cross section, summing
over struck protons and neutrons. The boundaries of the integration
domain $\Sigma$ in the $({\cal E},p)$ plane are found through the
energy conservation relation (see \cite{CDM,Barbaro:2008zv} for the
explicit expressions).

A further approximation underlies \eqref{eq:RL}, namely the factorization
of the single-nucleon response $R_L^{s.n.}(q,\omega)$ out
of the integral. Actually this response 
is in general half-off-shell and hence a function not only
of $q$ and $\omega$, but also of the energy and momentum of the
off-shell struck nucleon, or equivalently of $p$ and ${\cal E}$.
In the models being considered in the present study the struck
nucleon is in fact on-shell and so $R_L^{s.n.}$ becomes simply the
longitudinal response of a moving free nucleon. In this case its dependence
upon $p$ and ${\cal E}$ becomes very weak, particularly if one limits
the focus only to regions where the spectral function $S(p,{\cal
E})$ plays a significant role, and can accordingly be
extracted from the integral.

Finally, in this study we confine ourselves to dealing with
infinite, homogeneous systems, the simplest among them being the RFG
model in which the dynamics are controlled by just one parameter,
the Fermi momentum $k_F$. To explore superscaling it then turns out
to be convenient to recast \eqref{eq:RL} in the following form
\begin{equation}
\label{eq:RL1} 
R_L(q,\omega)={\cal N}\, R_L^{s.n.}(q,\omega)\, \Lambda
\times f(q,\omega)\, ,
\end{equation}
where
\begin{equation}
\label{eq:K} \Lambda =
\frac{1}{k_F}
\left(\frac{m_N}{q}\right)
\left(\frac{2 m_N T_F}{k_F^2}\right)
\simeq
\frac{m_N}{k_F q}~,
\end{equation}
$T_F$ being the Fermi kinetic energy.
The function
\begin{equation}
\label{eq:f} 
f(q,\omega)= 2\pi m_N k_F \times \frac{k_F^2}{2 m_N T_F}\,
\int\!\!\!\!\int_\Sigma dp\,d{\cal E}\,\frac{p}{E_p} S(p,{\cal E})\, ,
\end{equation}
is the so-called {\em superscaling function}.
Indeed, as we shall see in the next Section, in the RFG the function $f$
loses any dependence on both $k_F$ and $q$, namely, one has
superscaling in the non-Pauli-blocked regime. It remains
to be seen what happens in the BCS model, {\it i.e.,} in the
presence of correlations.


\section{The RFG model and its BCS-inspired extension}
\label{sec:model}

Before presenting our model for the correlated system, 
let us shortly recall the Fermi gas result.
The key tool for exploring
superscaling is the nuclear spectral function $S(p,{\cal E})$.
In the RFG model this reads~\cite{CDM,rely}
\begin{equation}
\label{eq:SRFG}
S^{RFG}(p,{\cal E}) = 4\, \theta(k_F-p) \delta({\cal
  E}-T_F+T_p) \frac{V_A}{A(2\pi)^3}~,
\end{equation}
where $T_p=E_p-m_N$ 
is the struck nucleon kinetic energy, $A$ the number of nucleons and $V_A$ 
the volume enclosing the system.
The integral \eqref{eq:f}
yields the RFG superscaling function~\cite{Alberico:1988bv}
\begin{equation}
\label{eq:fRFG} f_{RFG}(\psi) = \frac{3}{4} \left(1-\psi^2\right) \theta
\left(1-\psi^2\right)~,
\end{equation}
which depends only on one variable, defined as follows
\begin{equation}
\label{eq:psi} \psi = \frac{1}{\sqrt{\xi_F}}\times
\frac{\lambda-\tau}{\sqrt{(1+\lambda)
\tau+\kappa\sqrt{\tau(1+\tau)}}}~,
\end{equation}
with $\xi_F=T_F/m_N$, $\kappa=\frac{q}{2m_N}$, 
$\lambda=\frac{\omega}{2m_N}$ and $\tau=\kappa^2-\lambda^2$. 

As outlined in the Introduction, we now extend the RFG model in order to 
account for NN correlations by assuming for both the  
initial ground state ($|BCS>$) and the daughter nucleus ($|D(p)>$) 
a BCS-like wave function, namely
\begin{eqnarray}
\label{eq:BCS}
  |BCS>&=&\prod_k(u_k+v_k a^\dagger_{k\uparrow}a^\dagger_{-k\downarrow})|0>
\\
\label{eq:daugh}
  |D(p)>&=&\frac{1}{|\vp_p(p)|}\,a_{p\uparrow}
\prod_k[\up_k(p)+\vp_k(p)
   a^\dagger_{k\uparrow}a^\dagger_{ -k\downarrow}]|0>~.
\end{eqnarray}
In the above 
$|0>$ is the true vacuum and the states are correctly normalized providing
$|u_k|^2+|v_k|^2=1$ and $|\up_k(p)|^2+|\vp_k(p)|^2=1$.
Note that the $(u,v)$ and $(\up,\vp)$ coefficients are a priori different
from each other: this point is of crucial relevance for our
model, as we shall see below. 

With the assumption \eqref{eq:BCS} we have a covariant
approximation to the nuclear ground state wave function. We have
required that the added pairs always occur with back-to-back momenta
(hence the net linear momentum of the system in its rest frame is
zero) and with opposite helicities (hence the net spin of the
ground state is zero). The creation operators add particles with relativistic
on-shell spinors.

As is well-known, the states \eqref{eq:BCS} and \eqref{eq:daugh} do not correspond to a fixed 
number of particles, since they are not
eigenstates of the operator
$\hat n(k)=\sum_s a^\dagger_{k s}a_{k s}$.
However we can compute the expectation values
\begin{eqnarray}
\label{eq:nBCS}
  n_{BCS}(k)&=&<BCS|\hat n(k)|BCS>=|v_k|^2\\
  n_{D(p)}(k)&=&<D(p)|\hat n(k)|D(p)>
  =|\vp_k(p)|^2 (1-\delta_{kp})~.
\end{eqnarray}
and require the particle number ($A$ for the initial state and 
$A-1$ for the daughter nucleus) to be conserved on the average, which implies 
the conditions
\begin{eqnarray}
\label{eq:constr}
&&  \sum_k |v_k|^2 = A \,,
\,\,\,\,\,\,\,\,\,
\sum_{k\not=p}|\vp_k(p)|^2 = \sum_{k}|\vp_k(p)|^2-|\vp_p(p)|^2
=A-1~.
\end{eqnarray}

Concerning the energy, we view our system as being constructed in
terms of independent quasi-particles, writing accordingly
\begin{eqnarray}
\label{eq:EBCS}
  E_{BCS} &=& 
<BCS|\sum_{ks}E_k a^\dagger_{ks}a_{ks}|BCS>=
\sum_k E_k |v_k|^2\\
E_{D(p)} &=& <D(p)|\sum_{ks}E_k a^\dagger_{ks}a_{ks}|D(p)>
= \sum_{k\not=p} E_k |\vp_k(p)|^2
\nonumber\\
&=& (E_{BCS}-m_N)-T_p |v_p|^2+\sum_{k\not=p} T_k 
\left[|\vp_k(p)|^2-|v_k|^2\right]~,
  \label{eq:Edau}
\end{eqnarray}
where in the last equation the constraints (\ref{eq:constr}) have
been exploited.

Before computing the spectral function, let us write down the 
expressions for the normalization conditions (\ref{eq:constr}) 
in the thermodynamic limit
$A\to\infty$, $V_A\to\infty$, $A/V_A=\rho_A$, $(A-1)/V_{A-1}=\rho_{A-1}$,
namely
\begin{eqnarray}
&& \lim \frac{1}{V_A} \sum_k |v_k|^2 =
\int\frac{d^3k}{(2\pi)^3} |v(k)|^2 =\rho_A~,
\label{eq:norma}
\\
&& \lim \frac{1}{V_{A-1}}\left[\sum_k |\vp_k(p)|^2
-|\vp_p(p)|^2 \right]
= \int\frac{d^3k}{(2\pi)^3} |\vp(k;p)|^2 
= \rho_{A-1}~.
\label{eq:normb}
\end{eqnarray}
Assuming now $\rho_{A-1}=\rho_A\equiv\rho$ clearly entails
$\vpsq(k;p)=v^2(k)$, which allows us to drop the last term
in \eqref{eq:Edau}.  
It must be emphasized that the coefficients $v$ and $v^\prime$ become identical
in the thermodynamic limit, but are different for finite $A$. Hence
it is crucial to compute the nuclear energies when
$A$ is finite and {\em then} take the thermodynamic limit. 

We can then proceed to compute the daughter nucleus spectral
function
\begin{equation}
\label{eq:SFBCS}
  S^{BCS}(p,{\cal E})=\left|<D(p)|a_{p\uparrow}|BCS>\right|^2
\delta\left[{\cal E}-\left(E_{D(p)}-E_{D(k_F)}\right)\right]
\frac{V_A}{A(2\pi)^3}~,
\end{equation}
where $E_{D(k_F)}$ is the energy $E_{D(p)}$ of the daughter
nucleus evaluated at that value of $p$ where it reaches its
minimum, to be referred to as $k_F$ in the BCS model:
\begin{equation}
\label{eq:stab}
\left.\frac{d E_{D(p)}}{dp}\right|_{p=k_F}=0~.
\end{equation}
Hence we
have, after taking the thermodynamic limit,
\begin{equation}
{\cal E}(p) = E_{D(p)}-E_{D(k_F)} =
T_F |v(k_F)|^2 - T_p |v(p)|^2~.
\label{eq:cale}
\end{equation}
The matrix element in
\eqref{eq:SFBCS} can be straightforwardly computed, yielding $|v(p)|^2$,.
Thus we end up with the expression
\begin{equation}
  S^{BCS}(p,{\cal E}) =  |v(p)|^2 \delta\left({\cal E}-
T_F |v(k_F)|^2 + T_p |v(p)|^2\right)
\frac{1}{(2\pi)^3 \rho}~.
\label{eq:SFBCS1}
\end{equation}

Finally, in order to calculate the superscaling function \eqref{eq:f} 
what remains to be specified is the integration
region $\Sigma$, which in turn requires knowledge of the separation
energy. In the present model the latter turns out to be
$E_s = -T_F |v(k_F)|^2$.

The last ingredient needed to calculate $f$ are
the coefficients $v(k)$ appearing in the BCS wavefunction. Although in 
principle these could be computed self-consistently, together with the 
energies $E_k$, starting from a model Hamiltonian, here we take a more 
phenomenological approach, choosing the following three-parameter expression
\begin{equation}
  v^2(k)=\frac{c}{e^{\beta(k-\tilde k)}+1}~.
\label{eq:v2}
\end{equation}
Moreover, for sake of simplicity, we make the assumption
$E_k=\sqrt{k^2+m_N^2}$, namely we take the same single particle energies as in 
the RFG.

Next we use the constraints \eqref{eq:norma} and \eqref{eq:stab}
to fix the parameters $c$ and $\tilde k$,
obtaining
\begin{equation}
\label{eq:c1}
c(\beta,\tilde k) = -
\pi^2\beta^3\rho
/  Li_3\left(-e^{\beta\tilde k}\right)
\end{equation}
and
\begin{equation}
\label{eq:ktilde}
  \tilde k=k_F+\frac{1}{\beta}\log\left[\frac{\beta}{k_F}
\sqrt{k_F^2+m_N^2} \left(\sqrt{k_F^2+m_N^2}-m_N\right)-1\right]~.
\end{equation}

As far as the parameter $\beta$ is concerned, it clearly controls
both the modifications of the momentum distribution near the Fermi
surface (promotion of pairs due to residual NN interactions, both
long- and short-range) and also the tail of the momentum
distribution due to short-range NN correlations.
Indeed, for $\beta$ very large one
recovers the familiar $\theta$-distribution of the RFG, while for smaller and
smaller $\beta$ more and more particles are pulled out of the Fermi
sea and produce a significant tail for the momentum distribution at
large momenta. The impact of the physics expressed by the parameter
$\beta$ on the superscaling function is explored in the next
section.


\section{Results}
\label{sec:results}

In presenting the results obtained using our model it is 
convenient to start by displaying the behaviour of the parameters
$\tilde k$ and $c$, which are fixed by the physical conditions of
normalization and stability, versus $\beta$ for given $k_F$. When
$\tilde k$, $c$ and $\beta$ are known so are the wave functions of
the initial and final nuclei.

\begin{figure}
\label{fig:fig1}
\begin{center}
\includegraphics[scale=0.7]{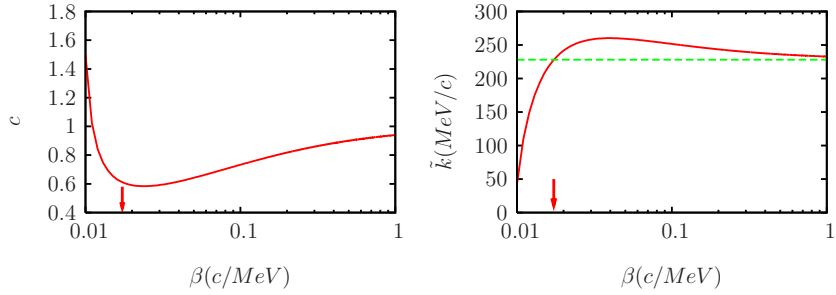}
\caption{The parameters $c$ and $\tilde k$, given in (\ref{eq:c1}) and
(\ref{eq:ktilde}), respectively,
shown as  functions of $\beta$ for $k_F=228$ MeV/c and $\rho=k_F^3/(6\pi^2)$.
The arrow indicates the critical value $\beta_{\rm crit}=0.017$ c/Mev and
the horizontal line in panel $a$ corresponds to the Fermi momentum $k_F$.}
\end{center}
\end{figure}
In Fig.~1 the parameters $\tilde k$ and $c$ are plotted
versus $\beta$. For large $\beta$ they stay constant 
(in fact the almost constant
value of $\tilde k$ is quite close to the input value $k_F=$ 228 MeV/c) until a critical value $\beta_{\rm
  crit}=0.017$ c/MeV is reached where $c$ ($\tilde k$) displays a dramatic
increase (decrease). This value corresponds to the change of sign
of the logarithmic term in \eqref{eq:ktilde}, namely
$\beta_{\rm crit}=\frac{2 k_F}{T_F (T_F+m_N)}$.
Thus our results appear to
point to the existence of a narrow domain of $\beta$ around
$\beta_{\rm crit}$, below which the system becomes strongly
disrupted by correlations. This has a strong impact on the structure
of the superscaling function, as we shall see later. 

In Fig.~2 we display the momentum distribution \eqref{eq:nBCS} of
the initial nucleus  for a few values of $\beta$ larger
(a) or smaller (b) than $\beta_{\rm crit}$. The
progressive development of a tail in the momentum distribution is
clearly seen in the figure:
for values of $\beta$ lower than $\beta_{\rm crit}$ the nuclear
momentum distribution becomes very much extended beyond the Fermi
sphere associated with the input value of $k_F$. 

\begin{figure}
\label{fig:fig3}
\begin{center}
\includegraphics[scale=0.7]{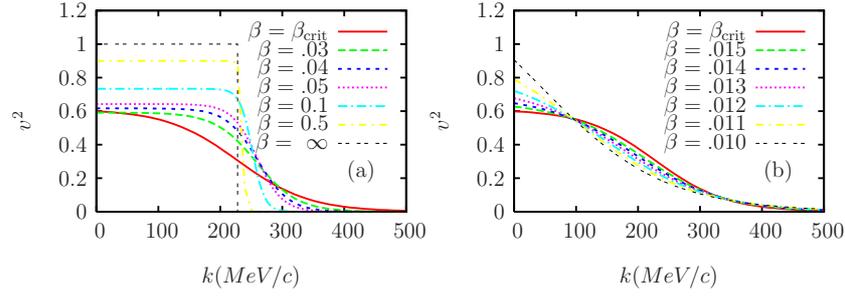}
\caption{Momentum distribution of the initial state, Eq.~(\ref{eq:v2}),
evaluated for $k_F=228$ MeV/c, $\rho=k_F^3/(6\pi^2)$ and
different values of $\beta$ (in c/MeV) above (a) and
below (b) the critical value $\beta_{\rm crit}$.}
\end{center}
\end{figure}

\begin{figure}
\label{fig:fig4}
\begin{center}
\includegraphics[scale=0.5]{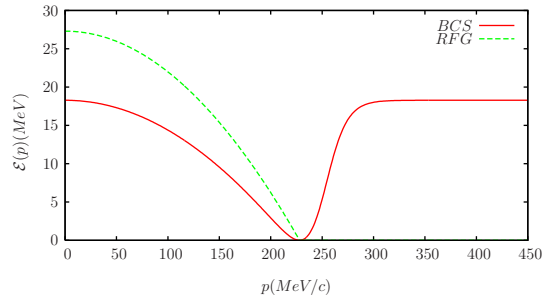}
\caption{The excitation energy $\cal E$ computed according to
\eqref{eq:cale} neglecting the last two terms for $k_F=228$
MeV/c, $\rho=k_F^3/(6\pi^2)$, $\beta=0.1$ c/MeV. The RFG results
are also shown for comparison. }
\end{center}
\end{figure}

The next issue to be addressed is to determine where the
spectral function is nonzero in the $({\cal E},p)$ plane. The
answer is found in Fig.~3 where the support of the spectral
functions of the RFG and of our BCS-inspired model are displayed
and compared. 
Both spectral functions of course are just
$\delta$-functions, but concerning their support two major
differences distinguish the two:
1) in the range of momenta where both exist the excitation
spectrum of the daughter system is substantially softer than the
RFG one; 2) for missing momenta larger than $k_F$ the BCS case,
unlike the RFG, continues to display a spectrum, which in the thermodynamic
limit rises quite suddenly with $p$ until it reaches the value
${\cal E}$ assumes for vanishing missing momentum, namely
${\cal E}_{\rm max} = T_F |v(k_F)|^2$.
This energy is reached only at
$p=\infty$, but over a large span of momenta ${\cal E}$ remains
almost constant, thus corresponding to the situation of an
eigenvalue with infinite degeneracy stemming from the symmetry
$U(1)$ associated with the particle number conservation. As $p$ is
lowered, approaching the Fermi surface, the degeneracy is lifted
and we face a situation of a spontaneously broken symmetry,
reflected in the structure of our state which contains components
of all possible particle number. This situation is strongly
reminiscent of superconductivity, where the spontaneous symmetry
breaking also occurs in the proximity of the Fermi surface.

\begin{figure}
\label{fig:fig5}
\begin{center}
\includegraphics[scale=0.8]{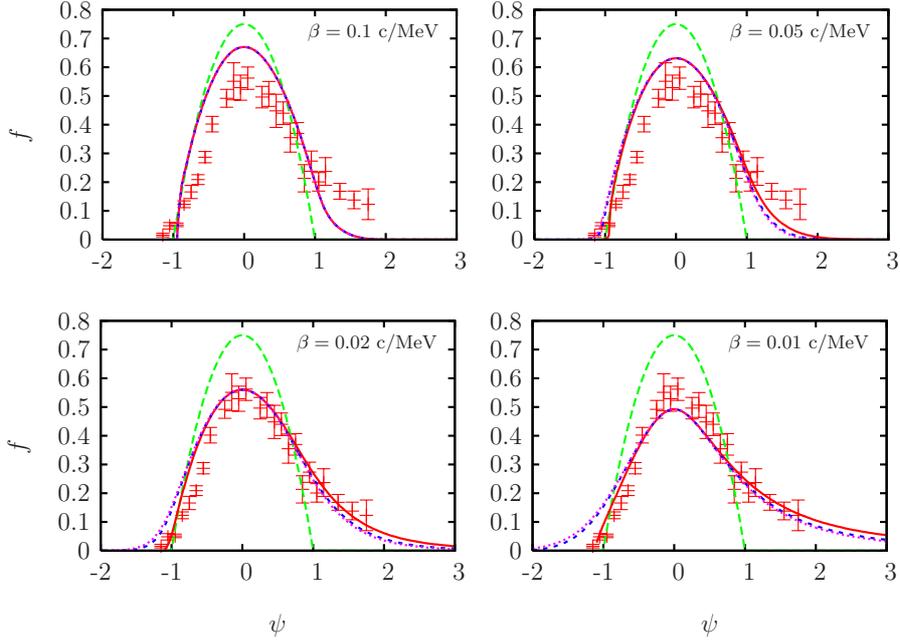}
\caption{The superscaling function $f$ defined in (\ref{eq:f})
plotted versus the scaling variable (\ref{eq:psi}) in the RFG model (green)
and in the present BCS model for three values of $q$ (red: 500 MeV/c,
blue: 1000 MeV/c, magenta: 1500 MeV/c) and
different values of $\beta$. As usual, $k_F=$228 MeV/c.
Data are taken from \cite{Maieron:2001it,JJ}.  }
\end{center}
\end{figure}
This set of degenerate states has a dramatic impact on the
superscaling function $f$, which is displayed in Fig.~4 versus the scaling
variable $\psi$  for a few values of $\beta$
and $q$. For comparison the RFG result in
\eqref{eq:fRFG} and the averaged experimental
data~\cite{Maieron:2001it,JJ} are also shown. One sees that to get
$f$ for large positive $\psi$ we have to integrate in the $({\cal
E},p)$ plane in domains encompassing large fractions of
those degenerate states discussed above. These are thus the cause of
the asymmetry of the scaling function with respect to $\psi=0$
appearing in Fig.~4. For $\psi$ large and negative these states are
to a large extent excluded from entering into the building up of
$f$. The fact that this effect is more and more pronounced as
$\beta$ becomes smaller reflects the impact of the tail of the
momentum distribution which indeed grows when $\beta$ decreases and,
as a consequence, more degenerate states participate to build up
$f$. Note that values of $\beta$ around the critical value
yield a tail which is in qualitative agreement with the experimental data.

\begin{figure}
\label{fig:fig6}
\begin{center}
\includegraphics[scale=0.7]{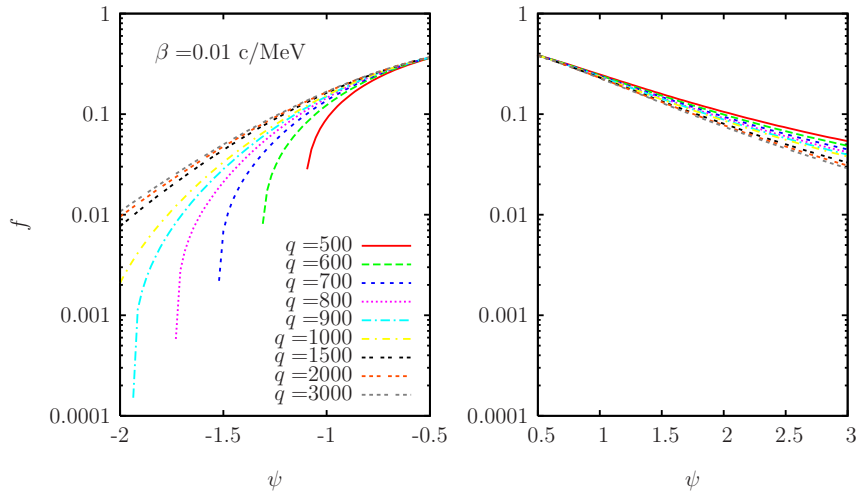}%
\caption{The superscaling function $f$ in the negative $\psi$
region plotted for several values of $q$ (in MeV/c) and
$\beta=0.01$ c/MeV. As usual, $k_F=$228 MeV/c.}
\end{center}
\end{figure}

As far as scaling of the first kind is concerned, Fig.~4 shows that
this is quickly reached in the vicinity of the QE peak, although not so
to the right and to the left of it. A closer examination of the
results (see Fig.~5, where $f$ is plotted on a logarithmic
scale for a wider $q$-range at $\beta$=0.01 c/MeV) shows
that also here the BCS model does scale, however with an onset
reached only for $q\simeq$ 1.5 GeV/c, namely for larger momenta than
when at the QE peak where the onset already occurs at about 500 MeV/c.
Also from Fig.~5 it appears that the scaling regime is
reached faster to the right than to the left of the QE peak. Moreover,
the asymptotic value for $\psi<0$ is approached from below, namely
the superscaling function grows with $q$ until it reaches its
asymptotic value, in contrast with the experimental findings. This
reflects the fact that our model, although appealingly simple, is
not able to account for features of this kind. 
Note that the same trend of
approaching first-kind scaling from below is also found
within the framework of the
Coherent Density Fluctuation Model\cite{Martin} 
where realistic nucleon momentum
and density distributions are used~\cite{bulgari}. On the other hand, in
relativistic mean-field theory~\cite{Caballero:2005sj} the approach
is from above, and thus in better accord with the experimental data.

Finally, using the present BCS model, we investigate the second-kind
scaling behaviour, namely the dependence of the function $f$ upon the nuclear 
species.
Following the original procedure of Refs.~\cite{DS},
we choose for each nuclear species a momentum $k_A$ 
(which is a phenomenological parameter, not necessarily coinciding with the
Fermi momentum as it must reflect both initial- and
final-state interaction effects) 
and use it in the definition \eqref{eq:psi} of the scaling
variable $\psi$ and of the dividing factor \eqref{eq:K}.
For simplicity, in the present approach the value of $k_A$ is chosen
in order to have all the corresponding superscaling functions
coincide at the QE peak, thus realizing superscaling at least where the
nuclear response is the largest. The results are displayed in Fig.~6,
where each curve corresponds to given $k_F$ and $k_A$. 
Over much of the range of $\psi$ shown in the figure
one sees relatively good second-kind scaling, although the results
still point to a sizable violation of the second kind scaling in the
scaling domain (large negative $\psi$).
\begin{figure}
\label{fig:fig9}
\begin{center}
\includegraphics[scale=0.7]{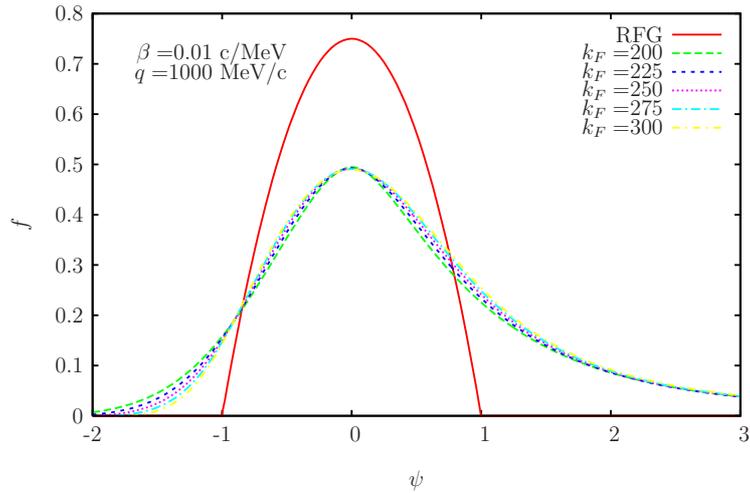}
\caption{The superscaling function $f$ plotted versus $\psi$ for
several values of the Fermi momentum $k_F$ (in MeV/c) and with $k_A$
devised in such a way that the peaks coincide (see text).}
\end{center}
\end{figure}


\section{Conclusions}
\label{sec:concl}

In the present study a simple extension of the relativistic Fermi
gas model for studies of relatively high-energy inclusive
electroweak cross sections has been developed. Starting from the RFG
in which a degenerate gas of nucleons is assumed for the nuclear
ground state, in this extension pairs of particles are promoted from
below the Fermi surface to above, yielding a spectral function and
the resulting momentum distribution with Fourier components for all
values of momentum. In the spirit of the RFG this new model has been
constructed in a way that maintains covariance.

To summarize our findings, we have shown that,
likely because in the BCS spirit we limit ourselves
to an independent quasi-particle description of nuclear matter,
scaling of the first kind (independence of momentum transfers $q$)
appears to occur not only at the QE peak, but also
at both lower and higher energy transfers $\omega$. We found that the
onset of first-kind scaling already occurs at momentum transfers of
order 500 MeV/c at the QE peak, whereas away from the QE peak the onset only occurs at
quite large momentum transfers (of the order of 2 GeV/c).
Furthermore, the shape of the superscaling function
turns out to be {\em non-symmetric} around the QE peak, being larger to
the right and smaller to the left of it, namely, in agreement with
experiment and thus lending support to our approach. However, in our
model when in the so-called scaling region (below the QE peak)
first-kind scaling is reached as a function of $q$ from below,
which is not what is
experimentally found. Finally, scaling of the second kind
(independence of nuclear species) is 
shown to be relatively well satisfied,
given that an appropriate momentum scale is chosen for each nuclear
species, although some violations appear for large negative $\psi$.



\end{document}